\documentclass[12pt,openright]{article}
\usepackage{graphicx}
\usepackage{amssymb}		
\usepackage{lmodern}
\usepackage[]{algorithmicx}
\usepackage{algpseudocode}
\usepackage{sectsty}
\usepackage{setspace}
\usepackage{lineno}
\usepackage{amsmath}
\usepackage{graphicx}

\sectionfont{\fontsize{13}{15}\selectfont}
\subsectionfont{\fontsize{11}{15}\selectfont}
\usepackage{geometry}
\begin{document}
\title{\textbf{\large Distances in and Layering of a DAG}
\date{\vspace{-5ex}}
\author{{\footnotesize $^{1,2}$Bhadrachalam Chitturi, $^1$Priyanshu Das }}}
\maketitle
\begin{center}
{\footnotesize $^1$Department of Computer Science and Engineering,  Amrita University, Amritapuri, India.}\\
{\footnotesize $^2$Department of Computer Science, University of Texas at Dallas, Richardson, Texas 75083, USA.}
\end{center}

{\footnotesize\textbf{Abstract:}}
The diameter of an undirected unweighted graph $G=(V,E)$ is the maximum value of the distance from any vertex $u$ to another vertex $v$ for $u,v \in V$ where distance i.e. $d(u,v)$ is the length of the shortest path from $u$ to $v$ in $G$. 
DAG, is a directed graph without a cycle. We denote the diameter of an unweighted DAG $G=(V,E)$ by $\delta (G)$.
The stretch of a DAG $G$ is the length of longest path from $u$ to $v$ in $G$, for all choices of $(u, v) \in V$ denoted by $\Delta (G)$. The diameter of an undirected graph can be computed  in $O(|V|(|V|+|E|))$ time by  executing breadth first search $|V|$ times. We show that stretch and diameter of a DAG can be computed in $O(|V|+|E|)$ time and $O(|V||E|)$ time respectively. 

A DAG is balanced if and only if a consistent assignment of level numbers to all vertices is possible. Layering refers to such an assignment. A balanced DAG is defined. An efficient algorithm that either detects whether a given DAG is unbalanced or layers it otherwise is designed with a running time of $O(|V|+|E|)$. 
\\
Key words: Diameter, directed acyclic graph, longest directed path, graph algorithms, complexity.
\section{Introduction}
A graph $G=(V,E)$  is a directed acyclic graph, i.e. $DAG$, if $G$ G is directed and has no  cycles. Further if edges do not have any weights then the DAG is unweighted. In this article we consider only unweighted DAG. The algorithms can be easily extended to a weighted DAG. The longest path problem from a given source vertex $s$ is hard to compute on an undirected graph; however, the same can be computed in a DAG in $O(|V|+|E|)$ time. 
The diameter in an undirected graph can be directly obtained  by executing a BFS from each node in $O(|V|(|V|+|E|))$ time \cite{cormen2011algorithms, sedgewick2011algorithms, kleinberg2006algorithm, brassard1988algorithmics, dasgupta2006algorithms}. 
We follow the standard terminology and call the maximum distance in a DAG $G$ as the diameter of $G$ denoted by $\delta(G)$. Likewise, the length of the longest path in a DAG is called its \emph{stretch} denoted by $\Delta(G)$. We show that stretch can be computed in linear time.


\section{Diameter and Stretch}
The following algorithm computes the \emph{stretch}, i.e. $\Delta(G)$ of G. It employs dynamic programming with memoization.\\\\
\textbf{Stretch Algorithm}
\begin{itemize}
\item {Let $lp(i)$ denote the maximum length of any path from $i$ and $LP(i)$ is the corresponding function call. Let $I$ be the set of vertices with no incoming edges. Note that $\forall_{i \in I}  max (LP(i))$ is the \emph{stretch} of G.  }
\item {Initialize: $\forall_{i \in V} lp[i] = -1$; (all vertices are unvisited).}
\item { Execute $\forall_{i \in I} LP(i)$.}
\item {$\delta(G) = max (\forall_{i \in I}~~lp[i] )$}
\end{itemize}
\begin{itemize}
\item{ LP(i) \{\\
if ($lp[i] \neq -1$) return ($lp[i]$);\\
\hspace{16mm} else \\
\hspace{8mm} $lp[i]=max(\forall_{u \in child[i]}~LP(u)+1)$\\
\hspace{8mm} return $lp[i]$;\\
            \}
    }
\end{itemize}

\textbf{Time Complexity}\\
Since every vertex $v$ is visited once, and $lp[v]$ is updated maximum as many times as the number of outgoing edges $v$ has, the overall complexity is $O(|V|+|E|)$\\

\noindent\textbf{Diameter Algorithm}\\
Let S ($\subset V$)) be set of vertices with no outgoing edges. For every vertex note the vertices from which it has an incoming edge. For every vertex $u \in S$, the set of vertices that $u$ can reach is $\phi$. Let $d(u,v)$ denote the distance from $u$ to $v$. 
\begin{itemize} 
\item{While S is not empty}
\begin{itemize} 

\item{For all edges $(p, v)$ where $v \in S$.\\
$d(p,v) \leftarrow 1$;  \\
if $x$ can be reached from $v_i$ where $v_i$ is an immediate successor of $p$ then\\  
$d(p,x) \leftarrow \forall_i~ ( min( d(v_i,x)+1) )$;
        }
\item{ Among the vertices that have an outgoing edge to some vertex in $S$ choose the vertex set $X$ in which all the vertices will have no outgoing edges once S is deleted.  $S \leftarrow X$}
\end{itemize} 
\end{itemize}

\noindent\textbf{Time Complexity}\\
For every vertex $u$ and $v$, where $v$ can be reached from $u$, $dist(u, v)$ needs to be updated $q$ number of times, where $q$ is the number of outgoing edges from $u$. Since, for every $u$, there can $O(|V|)$ vertices it can reach, and total number of edges is $E$, and each edge implies that the child updates the parent in $O(|V|)$ time, the total complexity is $O(|E|*|V|)$.
\section{Layering}
A DAG in which the lengths of all paths from $u$ to $v$ are identical is called as a \emph{balanced DAG}.
Labeling is a many to one function that assigns non-negative integers to the vertices such that the following conditions hold for any pair of vertices. The label of any vertex $v$ that has an incoming path length of $l$ from a vertex $u$ is $label(u)+l$ and the label of a vertex $v$ from which there is a path of length $l$ to  $u$ is $label(u)-l$. The idea of labeling is to assign consistent \emph{layer} numbers to the vertices. The layer indicates the relative depth of a vertex with respect to a vertex that has the lowest layer number.
The following algorithm labels a balanced DAG.\\

\noindent\textbf{Label Algorithm1}\\
We consider a DAG $G$ on which the stretch has been computed in linear time as shown earlier.
\begin{itemize}
\item  Initialize $\forall_{v \in V} label[v]=INV;$ where INV is an invalid value. Let $I$ be the set of vertices that do not have an outgoing edge.
\item Label one of the nodes say $i$ in $I$  where $lp[i]=\delta(G)$ with zero. That is, $label[i]=0$.
\item Note that such a vertex $i$ can be chosen in $O(|E|+|V|)$ time by the Stretch Algorithm.
\item We define a priority queue $Q$ where each object corresponding to a vertex stores an ordered pair of integers  $(label, u)$ where $label$ is the label number of the vertex $u$. 
\item $Q.push(0,i)$. 
\item Repeat while $Q \neq \phi$ 
\begin{itemize}
\item Let the object with lowest priority in $Q$ be $o$. Pop $o$ from $Q$.
\item If any parent $p$ of $o$ is not labeled then\\ $label[p]= label[o]-1$; $Q.push(label[p],p)$; 
\item If any child $c$ of $o$ is not labeled then\\ $label[c]= label[o]+1$; $Q.push(label[p],p)$;
\end{itemize}

\item Let $m$ be $min (\forall_{i \in V}~~label[i] )$ $\forall_{u \in V}~~label[v] = label[v] + \mid m \mid$\\
\end{itemize}

\noindent\textbf{Analysis}\\
The first time a vertex $v$ is visited, it is assigned a label.
If a label has been assigned to a vertex $a$  hen we know its distance from any vertex $b$ where $a$ can be reached from $b$ and $b$ is labeled. 
At the end of the algorithm all labels are updated such that they are non-negative integers. A final label is the \emph{layer} number of the corresponding vertex. Thus, we call this process as \emph{layering}. \\

\noindent\textbf{Time Complexity}\\
Each edge is processed only once, and every vertex is pushed into $Q$ only once. Therefore, accessing all vertices in $Q$ takes $O(|V|log|V|)$ time. Thus, time complexity of the labeling, given that the given graph is \emph{balanced DAG} is $ O(|E|+|V|log|V|)$. \\

\noindent\textbf{Detection of Unbalanced DAG}\\
If we apply the above algorithm to label a DAG. At any point, if we try to label a vertex which is already labeled, and the new label conflicts with the old label, we know that there exists at least one pair $(u, v)$ such that there are two paths from $u$ to $v$ with different path lengths. Time complexity of detection, therefore, is the same as the time complexity of labeling, i.e. $O(|E|+|V|log|V|)$.\\

\noindent\textbf{Label Algorithm2}\\
The input to this algorithm is a DAG $G$ on which the stretch has been computed. Let $I$ be the set of vertices that do not have an outgoing edge.
\begin{itemize}
\item  Initialize $\forall_{v \in V} label[v]=INV;$ where INV is an invalid value.
\item Label one of the nodes say $i$ in $I$  where $lp[i]=\delta(G)$ with zero. That is, $label[i]=0$.
\item Note that such a vertex $i$ can be chosen in $O(|E|+|V|)$ time by the Stretch Algorithm.
\item {Define $Label(v, l)$\\
    $label[v] = l$\\
    $\forall_p \exists (p, v) \in E$\\
        if $label[p] == INV$\\
            $Label(p, l-1)$\\
    $\forall_c \exists (v, c) \in E$\\
        if $label[c] == INV$\\
            $Label(c, l+1)$\\
}
\item {Call $Label(i, 0)$}
\item Let $m$ be $min (\forall_{i \in V}~~label[i] )$ $\forall_{u \in V}~~label[v] = label[v] + \mid m \mid$\\
\end{itemize}

\noindent\textbf{Analysis}\\
The first time a vertex $v$ is accessed, it is assigned a label. If at any point, a label is assigned to a vertex $v$, we know it's distance from any vertex $u$ where $v$ can be reached by $u$ and $u$ is labeled. Just before the algorithm terminates, all labels are updated such that they are non-negative integers. \\

\noindent\textbf{Time Complexity}\\
The Label function is called for every node only once, which means $O(|V|)$ function calls. Also, every function call from a vertex $u$ examines its immediate successors' and immediate descendants' labels. So, the amortized time complexity for accessing labels = $O(|E|)$. Therefore total time complexity is $O(|V|+|E|)$.\\

\noindent\textbf{Detection of Unbalanced DAG}\\
Let the above algorithm be run to label a DAG. If a vertex which is already labeled is attempted to be labelled again and the new label conflicts with the existing label then the DAG is unbalanced. This is so because there exists at least one pair $(u, v)$ such that there are two paths from $u$ to $v$ with different path lengths. Time complexity of detection, therefore, is the same as the time complexity of labeling, i.e. $O(|E|+|V|)$.

\section{Conclusion}
We show that $\Delta (G)$ and $\delta(G)$ of DAG can be computed in $O(|V|+|E|)$ time and $O(|V||E|)$ time respectively. Further, we define balanced DAG and design an efficient algorithm to layer a balanced DAG. A DAG that is not balanced does not have a well defined layering.
One can either layer a balanced DAG or determine that the given DAG is unbalanced in  $O(|V|+|E|)$ time.

{}


\begin{thebibliography}{}
\bibitem{cormen2011algorithms}
Cormen, Thomas H. and Leiserson, Charles E. and  Rivest,  Ronald L. and Stein,  Clifford.
Introduction to Algorithms, MIT Press, 2011.

\bibitem{sedgewick2011algorithms}
Sedgewick, Robert and Wayne, Kevin.
Algorithms, Addison-Wesley Professional, 2011.

\bibitem{kleinberg2006algorithm}
Kleinberg, Jon and Tardos, Eva.
Algorithm design, Pearson Education India, 2006.

\bibitem{brassard1988algorithmics}
Brassard, Gilles and Bratley, Paul. Algorithmics: theory \& practice. Prentice-Hall, Inc.1988.

\bibitem{dasgupta2006algorithms}
Dasgupta, Sanjoy and Papadimitriou, Christos H and Vazirani, Umesh. Algorithms, McGraw-Hill, Inc., 2006.

\end{thebibliography}
\end{document}